\documentclass[twocolumn,showpacs,floats,floatfix,superscriptaddress,aps,pra]{revtex4}
\usepackage{amssymb}
\usepackage{amsmath}
\usepackage{graphicx}
\usepackage{bm}
\usepackage{bbm}

\usepackage[usenames]{color}

\def\be{ \begin{equation} }
\def\ee{ \end{equation} }
\def\bea{ \begin{eqnarray} }
\def\eea{ \end{eqnarray} }
\def\bse{ \begin{subequations} }
\def\ese{ \end{subequations} }

\def\half{\tfrac12}

\begin{document}

\author{I. E. Linington}
\affiliation{Department of Physics, Sofia University, James Bourchier 5 blvd, 1164 Sofia,
Bulgaria}
\affiliation{Department of Physics and Astronomy, University of Sussex, Falmer, Brighton,
BN1 9QH, United Kingdom}
\author{P. A. Ivanov}
\affiliation{Department of Physics, Sofia University, James Bourchier 5 blvd, 1164 Sofia,
Bulgaria}
\affiliation{QOLS, Blackett Laboratory, Imperial College London, SW7 2BW, United Kingdom}
\affiliation{Institute for Mathematical Sciences, Imperial College London, SW7 2PG,
United Kingdom}
\author{N. V. Vitanov}
\affiliation{Department of Physics, Sofia University, James Bourchier 5 blvd, 1164 Sofia,
Bulgaria}
\affiliation{Institute of Solid State Physics, Bulgarian Academy of Sciences,
Tsarigradsko chauss\'{e}e 72, 1784 Sofia, Bulgaria}
\author{M. B. Plenio}
\affiliation{QOLS, Blackett Laboratory, Imperial College London, SW7 2BW, United Kingdom}
\affiliation{Institute for Mathematical Sciences, Imperial College London, SW7 2PG,
United Kingdom}

\title{Robust control of quantized motional states of a chain \\of trapped ions by collective adiabatic passage}
\date{\today }

\begin{abstract}
A simple technique for robust generation of vibrational Fock states in a chain of trapped ions is proposed. 
The method is fast and easy to implement, since only a single chirped laser pulse, simultaneously addressing all of the ions, is required. 
Furthermore, because the approach uses collective adiabatic passage, significant fluctuations in the intensity or frequency of the laser pulse can be tolerated, and the technique performs well even on the border of the Lamb-Dicke regime. 
We also demonstrate how this technique may be extended in order to create non-classical superposition states of the ions' collective motion and
 Greenberger-Horne-Zeilinger states of their internal states.
Because only a single laser pulse is required, heating effects arising under realistic experimental conditions are negligibly small. 
\end{abstract}

\maketitle

\section{Introduction}
\label{introduction} 
The interaction of a single quantum harmonic oscillator
with one or more two-level systems is a fundamentally important model in
quantum physics \cite{JC} and offers an ideal arena for investigating the
principles of superposition and entanglement in mesoscopic systems 
\cite{ShoreKnight}. Many physical systems are well approximated by such a model,
including, for example, atom-cavity systems \cite{Raimond}, chains of
trapped ions with coupled internal and vibrational degrees of freedom 
\cite{Leibfried}, quantum dots interacting with photonic crystal fibres or
nanocavities \cite{Yoshie} and Josephson junctions interacting with
microwave-resonators \cite{Wallraff}. An area of active interest is to 
engineer non-trivial quantum states of the harmonic oscillator through a 
controlled manipulation of the two-level systems with which it interacts. 
This approach is well-established within cavity quantum electrodynamics 
where atoms passing through a cavity have been used to generate 
Schr\"odinger-cat states \cite{Meunier} and Fock states \cite{Varcoe} of 
the cavity field.

The motional state of one or more trapped ions is another highly promising
system in which to investigate non-classical harmonic oscillator states.
When a chain of ions is cooled sufficiently, each vibrational normal mode
closely approximates an ideal harmonic oscillator. By using appropriately
tuned lasers, the transitions in the internal state of each ion are
accompanied by changes in the vibrational state of a particular normal mode 
\cite{James, Leibfried}. Furthermore, each mode couples very weakly to its
environment (via fluctuating charges on the trapping electrodes), so
decoherence rates are typically much lower than in cavity QED systems.

Many theoretical proposals exist for the creation of non-classical harmonic
oscillator states in the motion of trapped-ions. In particular, the
preparation of motional Fock states (eigenstates of the harmonic oscillator)
has been proposed via the observation of quantum jumps \cite{CBPZ, Eschner},
utilising adiabatic passage techniques \cite{CBZ}, or using the method of
trapping states \cite{BCZ, Wallentowicz}. Several non-classical states have
been prepared experimentally and fully characterized by quantum state
tomography, including Fock states, superpositions of Fock states,
Schr\"odinger cat states and squeezed vacuum states \cite{Meekhof, Monroe, leibfried1996, leibfried1998}.
Furthermore, the decoherence of these states has been investigated by using
specially engineered reservoirs \cite{Myatt, Turchette}. However, despite
several promising theoretical proposals for their generation, to date,
motional Fock states have only been successfully created by applying a
sequence of Rabi $\pi$-pulses -- tuned to the blue- and red-sidebands for a
single ion \cite{Meekhof, Roos1999}. While this approach is conceptually
simple, it is also technically demanding, since an $N$-phonon Fock state
requires $N$ separate pulses of precise area and frequency, and each of 
these must be applied using a tightly-focussed beam addressing a single 
ion. Intensity and frequency fluctuations result in a reduced fidelity.

In this paper, we propose a novel and very simple adiabatic passage technique for the generation of Fock states, requiring only one common laser pulse addressing the whole ion chain. 
This technique can readily be adapted to the creation of Greenberger-Horne-Zeilinger (GHZ) states and superpositions of phonon Fock states. 
The approach has several significant advantages over existing proposals, namely: 
 (i) it is simple to implement, since only a single laser pulse is required;
 (ii) the state preparation is very fast, since it occurs in a single step;
 (iii) because there is only a very limited time for decoherence, the overall fidelity can be high;
(iv) our technique is naturally robust against fluctuations in the laser Rabi frequency and detuning, since it utilises adiabatic passage; (v) the technique holds up even when the first-order Lamb-Dicke approximation begins to break down.

The paper is arranged as follows. 
In Sec. \ref{Sec-DP}, we introduce a model Hamiltonian and Hilbert space for the case of $N$ trapped ions interacting with a common laser pulse. 
It is shown that by a judicious choice of basis (calculated using the generalized Morris-Shore transformation \cite{AVS}),
 the dynamics can be confined to an $(N+1)$-dimensional subspace of the overall ($2^{N}$-dimensional) Hilbert-space. 
Furthermore, by sweeping the laser detuning through resonance with a motional sideband, the ground state can be connected adiabatically to the highest energy state which has exactly $N$ vibrational quanta. 
This is shown explicitly for the first blue-sideband in Sec. \ref{create_Fock} and the first red-sideband in Sec. \ref{red_sideband}, using a `complex-sech' laser pulse shape. 
In Sec. \ref{create_GHZ}, it is shown that our technique may also be adapted in order to engineer GHZ states and superpositions of phonon Fock states of the ion chain. 
In Sec. \ref{heating} we estimate vibrational heating effects relevant to an experimental implementation of our technique.
Finally, in Sec. \ref{conclusions}, we conclude our findings.

\section{Definition of the problem \label{Sec-DP}}

We consider $N$ identical two-state ions confined in a linear trap and interacting with a common laser field. We assume that the ions are initially cooled to their ground state of motion \cite{King} and that the trapping frequencies are suitably chosen so that radial modes remain unexcited. 
The laser frequency $\omega _{L}(t)$ is tuned near to the centre-of-mass \textit{blue-sideband} resonance, $\omega _{L}(t)=\omega _{0}+\nu -\delta (t) $,
 where $\omega _{0}$ is the Bohr frequency of the transition, $\nu $ is the trap frequency and $\delta \left(t\right) $ is a time-dependent detuning. The chain of ions possesses $N$ axial vibrational degrees of freedom and the corresponding vibrational normal mode frequencies are denoted $\nu_{p}$ (where $p=1\ldots{}N$ and $\nu_{1}\equiv\nu$).
If the laser field has equal intensity at the position of each ion -- for example if it is directed along the trap axis -- then, after applying the electric-dipole and optical rotating-wave approximations, the interaction between the ions and the laser field is described by the following Hamiltonian \cite{James, buzek}:
\begin{align}
\mathbf{H}_{I}(t) = & \frac{\hbar\Omega(t)}{2}\sum_{j=1}^{N}\Bigg\{
\sigma _{j}^{+}\exp\left(i\int_{t_{i}}^{t}\delta\left(\tau\right)d\tau-i\nu{}t-i\phi _{j}\right) 
\nonumber \\
&\times{}\exp\left(\sum_{p=1}^{N}i\eta\kappa_{p}^{j}\left[a_{p}e^{-i\nu_{p}t} + a_{p}^{\dagger}e^{i\nu_{p}t}\right]\right)
+ h.c.\Bigg\}.
\label{Hamiltonian_full}
\end{align}
Here $\sigma _{j}^{+}=\left\vert 1_{j}\right\rangle \left\langle 0_{j}\right\vert $ and $\sigma _{j}^{-}=\left\vert 0_{j}\right\rangle \left\langle1_{j}\right\vert$ are the raising and lowering
operators for the internal states of the $j^{th}$ ion, $a_{p}^{\dagger }$ and $a_{p}$ are respectively the creation and annihilation operators for phonons in the $p^{th}$ mode. \mbox{$\eta =\sqrt{\hbar k^{2}\cos
^{2}\theta /2M\nu}$} is the single-ion Lamb-Dicke parameter, with $k$ being the laser
wavenumber, $\theta $ the angle between the trap axis and the direction of
the laser beam and $M$ the mass of the ion. $\kappa_{p}^{j}$ is a dimensionless coupling constant that depends both on the particular ion and the mode under consideration.
The function $\Omega(t)$ is the (real-valued) time-dependent Rabi frequency and
\begin{align}
\phi_{j}=\phi_{L} -kx_{j}^{0}\cos\theta
\label{phi_def}
\end{align}
is the phase of the laser pulse at the equilibrium position $x_{j}^{0}$ of the $j^{th}$ ion. 

\begin{figure}[tbp]
\includegraphics[width=55mm]{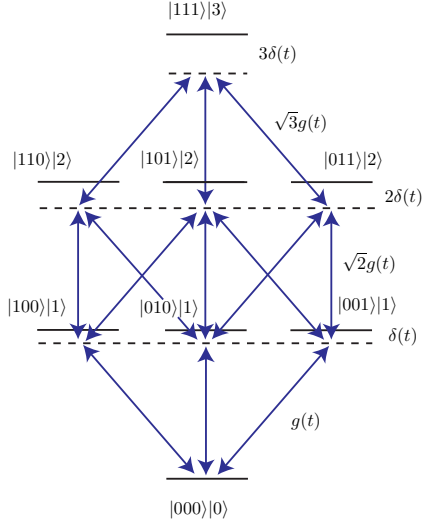}
\caption{(Color online.) Linkage pattern of the collective states of three trapped ions coupled to their common centre-of-mass mode. 
The difference between the number of ionic excitations and the number of phonons is conserved. 
The laser beam is tuned near to the blue-sideband resonance with detuning $\protect\delta \left( t\right) $. The coupling between the manifolds indexed by $n$ and $n+1$ is $g\left( t\right) \protect\sqrt{n+1}$.}
\label{Fig-1}
\end{figure}

The laser parameters are chosen to satisfy $\delta(t)/\nu\ll1$, $\Omega(t)<\nu$ and in this limit, off-resonant transitions to higher-order vibrational sidebands have a negligible effect on the dynamics. Therefore, to good approximation, we may write:
\begin{align}
\mathbf{H}_{I}(t) \approx & \frac{\hbar\Omega(t)}{2}\sum_{j=1}^{N}\Bigg\{
\sigma _{j}^{+}\exp\left(i\int_{t_{i}}^{t}\delta\left(\tau\right)d\tau -i\nu{}t -i\phi _{j}\right) 
\nonumber \\
&\times{}\exp\left(i\frac{\eta}{\sqrt{N}}\left[a{}e^{-i\nu{}t} + a^{\dagger}e^{i\nu{}t}\right]\right)
+ h.c.\Bigg\},
\label{Hamiltonian_COM}
\end{align}
where for notational convenience, we write \mbox{$a\equiv{}a_{1}$}, \mbox{$a^{\dagger}\equiv{}a_{1}^{\dagger}$} and we have used the fact that \mbox{$\kappa_{1}^{j}=1/\sqrt{N}$ $\forall{}\;j$} \cite{James}. 
Equation (\ref{Hamiltonian_COM}) simplifies considerably when $\eta\sqrt{n+1}\ll1$, where $n$ is the number of phonons in the centre-of-mass mode -- a condition known as the Lamb-Dicke limit. After expanding in powers of $\eta$ and applying the vibrational rotating-wave approximation, the dominant contribution to Eq. (\ref{Hamiltonian_COM}) is \cite{CZ1995} 
\begin{eqnarray}
\mathbf{H}_{I}\left( t\right) \approx \hbar g\left( t\right) \sum_{j=1}^{N}\left[
a^{\dagger }\sigma _{j}^{+}e^{i\int_{t_{i}}^{t}\delta \left( \tau \right)
d\tau -i\phi _{j}+i\pi/2} + h.c. \right],
\label{Hamiltonian}
\end{eqnarray}
where we have introduced a scaled coupling, \mbox{$g\left( t\right)=\eta \Omega \left( t\right) /2\sqrt{N}$}, between the internal and motional degrees of freedom. 
We remark that although the following analysis uses the simplified Hamiltonian (\ref{Hamiltonian}), the numerical simulations presented below are based on the Hamiltonian given in Eq. (\ref{Hamiltonian_COM}).  The effect of the additional counter-rotating terms and terms higher-order in $\eta$ is investigated in Sec. \ref{LD} and found to be negligible for a wide range of parameters.
The above Hamiltonian describes the anti-Jaynes-Cummings model, which conserves the
difference between the number of ionic and vibrational excitations -- a feature characteristic of blue-sideband excitation.
We assume that the initial state is $\left\vert 0_{1}0_{2}\ldots 0_{N}\right\rangle \left\vert 0\right\rangle $,
 where $\left\vert n\right\rangle $ is the Fock state with $n$ phonons ($n=0,1,2\ldots $) and $\left\vert 0_{1}0_{2}\ldots 0_{N}\right\rangle$ is the collective ground state of the ions. 
Each manifold with $n$ phonons and $n$ ionic excitations is \mbox{$C_{n}^{N}$-fold} degenerate, where \mbox{$C_{n}^{N}=N!/[n!( N-n)!]$}. 
An example linkage pattern for three ions is illustrated in Fig. \ref{Fig-1}.

%
The entire Hilbert space $\mathcal{H}$ is spanned by the following basis states: 
\begin{subequations}
\label{HS}
\begin{eqnarray}
&&\left\vert 0_{1}0_{2}\ldots 0_{N}\right\rangle \left\vert 0\right\rangle,
\label{ground} \\
&&\left\vert 0_{1}\ldots 1_{k}\ldots 0_{N}\right\rangle \left\vert
1\right\rangle,  \label{1exc} \\
&&\left\vert 0_{1}\ldots 1_{k}\ldots 1_{m}\ldots 0_{N}\right\rangle
\left\vert 0\right\rangle,  
\label{2exc} \\
&&\vdots  \notag \\
&&\left\vert 1_{1}1_{2}\ldots 1_{N}\right\rangle \left\vert N\right\rangle ,
\label{Nexc}
\end{eqnarray}
\end{subequations}
and has dimension $\dim \mathcal{H}=\sum_{n=0}^{N}C_{n}^{N}=2^{N}$.
We now perform a time-dependent phase transformation
\begin{equation}
\mathbf{U}(t)=\exp \left[ i\sum_{j=1}^{N}\left( \int_{t_{i}}^{t}\delta (\tau)d\tau -\phi _{j} + \pi/2 \right) \vert1_{j}\rangle\langle 1_{j}\vert\right] , 
\end{equation}
and re-express the Hamiltonian (\ref{Hamiltonian}) as
\begin{equation}
\mathbf{H}_{I}\left( t\right) =\mathbf{U}(t)\mathbf{\tilde{H}}_{I}\left(
t\right) \mathbf{U}^{\dagger }(t)+i\hbar \frac{\partial \mathbf{U}(t)}{\partial
t}\mathbf{U}^{\dagger }(t), 
\end{equation}
so that in the basis defined in (\ref{HS}), $\mathbf{\tilde{H}}_{I}\left(t\right) $ has a tridiagonal block-matrix form
\begin{equation}
\mathbf{\tilde{H}}_{I}\left( t\right) =\hbar \left[ 
\begin{array}{cccccc}
\mathbf{D}_{0} & \mathbf{V}_{0,1} & 0 & \ldots & 0 & 0 \\ 
\mathbf{V}_{0,1}^{\dagger } & \mathbf{D}_{1} & \mathbf{V}_{1,2} & \ldots & 0
& 0 \\ 
0 & \mathbf{V}_{1,2}^{\dagger } & \mathbf{D}_{2} & \ldots & 0 & 0 \\ 
\vdots & \vdots & \vdots & \ddots & \vdots & \vdots \\ 
0 & 0 & 0 & \ldots & \mathbf{D}_{N-1} & \mathbf{V}_{N-1,N} \\ 
0 & 0 & 0 & \ldots & \mathbf{V}_{N-1,N}^{\dagger } & \mathbf{D}_{N}%
\end{array}
\right] .  \label{HTDF}
\end{equation}
Here $\mathbf{V}_{n,n+1}\left( t\right) $ is a $\left( C_{n}^{N}\times C_{n+1}^{N}\right) $-dimensional interaction matrix which couples the $n$- and $(n+1)$-phonon manifolds. 
$(N-n)$ of the elements in each row and $(n+1)$ of the elements in each column of $\mathbf{V}_{n,n+1}\left( t\right)$ are non-zero, and from the symmetry of Eq. (\ref{Hamiltonian}),
 it can be seen that all of these non-zero elements are equal to $g(t)\sqrt{n+1}$.
$\mathbf{D}_{n}$ is a $C_{n}^{N}$-dimensional
diagonal matrix whose elements are equal to the respective shared detuning
of the $n$-phonon manifold 
\begin{equation}
\mathbf{D}_{n}\left( t\right) =n\delta \left( t\right) \mathbf{\mathbbm{1}}_{n},
\label{Delta}
\end{equation}
where $\mathbf{\mathbbm{1}}_{n}$ is the unit matrix. 

\begin{figure}[tbp]
\includegraphics[width=49mm]{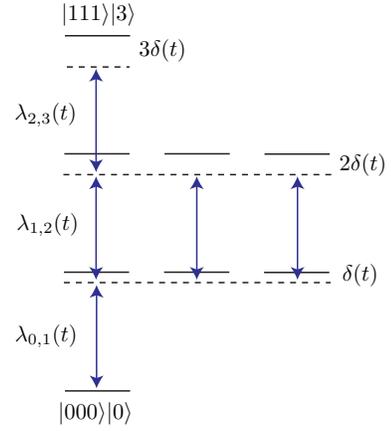}
\caption{(Color online.) Example of the Morris-Shore transformation for three ions. 
For the generation of Fock states, the dynamics is confined to a four-state chain, where the detuning of level $n$ is $n\delta(t)$ and the coupling between levels $n$ and $(n+1)$ is $\lambda_{n,n+1}(t)$.}
\label{Fig-2}
\end{figure}

On first inspection, an analytic treatment of the dynamics may appear to be highly demanding for more than a few ions, since in the original basis (\ref{HS}),
 the dynamics is described by $2^{N}$ coupled differential equations.
Fortunately however, by utilising the multilevel Morris-Shore transformation \cite{AVS}, it is possible to decouple this vast Hilbert space into a set of smaller closed subspaces,
 which simplifies the problem dramatically. 
Such a multilevel Morris-Shore factorisation exists if and only if the following commutation relations are satisfied for all pairs of neighbouring coupling-blocks $\left( n=1,\ldots ,N-1\right) $
\begin{equation}
\left[ \mathbf{V}_{n-1,n}^{\dagger }\mathbf{V}_{n-1,n},\mathbf{V}_{n,n+1}%
\mathbf{V}_{n,n+1}^{\dagger }\right] =0.  \label{C}
\end{equation}
It was recently shown that these commutation relations are automatically satisfied by the Hamiltonian given in Eq. (\ref{HTDF}) \cite{LV};
 the \mbox{$2^{N}$-dimensional} Hilbert space can therefore be factorized into a set of independent chainwise linkages, as shown in Fig. \ref{Fig-2}. 
Furthermore, for an appropriately chosen initial state, the dynamics is wholly confined within the largest of these subspaces -- an $(N+1)$-level ladder. 
The states which make up this chain are equal to 
\begin{align}
\mathbf{U}(t)\left\vert W_{n}^{N}\right\rangle \left\vert n\right\rangle \hspace{5mm}(n=0,\ldots ,N), 
\label{MS_states}
\end{align}
where the symmetric Dicke states $\left\vert W_{n}^{N}\right\rangle$ are defined as follows \cite{Dicke, MW} 
\begin{align}
\left\vert W_{n}^{N}\right\rangle =\frac{1}{\sqrt{%
C_{n}^{N}}}\sum_{k}P_{k}|\underbrace{1\ldots 1}_{n}\underbrace{0\ldots 0}%
_{N-n}\rangle,  
\label{DS}
\end{align}
and $\left\{ P_{k}\right\} $ above denotes all possible distinct permutations of the $n$ ionic excitations. 
A full derivation of the Morris-Shore factorisation described above can be found in \cite{LV}, but as a further justification,
 we note that $\mathbf{\tilde{H}}_{I}(t)$ is symmetric with respect to the interchange of any two ions. 
Therefore, states which are unchanged by a permutation of the ions can only be coupled by the Hamiltonian (\ref{HTDF}) to other states with this same symmetry. 
The initial and final states are both symmetric product states and therefore, the system begins and remains in the subspace defined by Eq. (\ref{MS_states}) throughout the dynamics.

When expressed in the factorized Morris-Shore basis, the entire evolution of the system is thus confined to an $(N+1)$-level ladder,
 where the lowest state is \mbox{$\left\vert0_{1}0_{2}\ldots 0_{N}\right\rangle \left\vert 0\right\rangle$},
 the highest is \mbox{$\left\vert 1_{1}1_{2}\ldots 1_{N}\right\rangle \left\vert N\right\rangle$},
 and all intermediate levels \mbox{($n=1,\ldots,N-1$)} are the multipartite entangled states (\ref{MS_states}). 
Each of the basis states contributing to level $n$ is connected to $(N-n)$ of the basis states which make up level $(n+1)$ in the ladder, and this coupling has strength $g(t)\sqrt{n+1}$. 
After a short calculation, the coupling between adjacent levels in the Morris-Shore chain is found to be
\begin{equation}
\lambda _{n,n+1}\left( t\right) =g\left( t\right) \left( n+1\right) \sqrt{N-n}.  
\label{CMS}
\end{equation}
Within the closed subspace formed by the $(N+1)$-level Morris-Shore ladder, the Hamiltonian reduces to 
\begin{equation}
\mathbf{H}_{N+1}\left( t\right) =\hbar \left[ 
\begin{array}{cccccc}
0 & \lambda _{0,1} & 0 & \ldots & 0 & 0 \\ 
\lambda _{0,1} & \delta & \lambda _{1,2} & \ldots & 0 & 0 \\ 
0 & \lambda _{1,2} & 2\delta & \ldots & 0 & 0 \\ 
\vdots & \vdots & \vdots & \ddots & \vdots & \vdots \\ 
0 & 0 & 0 & \ldots & \left( N-1\right) \delta & \lambda _{N-1,N} \\ 
0 & 0 & 0 & \ldots & \lambda _{N-1,N} & N\delta%
\end{array}%
\right] .  \label{HN+1}
\end{equation}%
As an alternative to the Morris-Shore formalism, we note that the state-space factorisation presented above can also be conveniently derived via a careful inspection of the Hamiltonian. A convenient route is to rewrite (\ref{Hamiltonian}) in terms of the combined ionic pseudo-spin operators, \mbox{$J_{z}=\half\sum_{j=1}^{N}\left(\vert{}1_{j}\rangle\langle1_{j}\vert - \vert0_{j}\rangle\langle0_{j}\vert\right)$}, \mbox{$J_{+} = \sum_{j=1}^{N}\vert1_{j}\rangle\langle0_{j}\vert$}, \mbox{$J_{-}=\sum_{j=1}^{N}\vert0_{j}\rangle\langle1_{j}\vert$}, as in Refs. \cite{molmer1999, unanyan2003, lopez2007}. Our approach complements the collective-spin picture, and is also more general, in that it is not strictly limited to the case of equal couplings \cite{AVS}.
%
\section{Creation of motional Fock states}\label{section_Fock}

\subsection{Creation of Fock states via a level crossing on the blue-sideband}\label{create_Fock} 

\subsubsection{Bow-tie ladder model}

We wish to transfer the population of the initial state $\left\vert 0_{1}0_{2}\ldots 0_{N}\right\rangle \left\vert 0\right\rangle$ to state $\left\vert 1_{1}1_{2}\ldots 1_{N}\right\rangle \left\vert N\right\rangle $. 
This can be done very efficiently by creating a `bow-tie' level crossing linkage pattern, wherein all energies [the diagonal elements of the Hamiltonian \eqref{HN+1}] cross at the same instant of time,
 and enforcing adiabatic evolution.
The convenience of the bow-tie linkage pattern for our multistate Morris-Shore ladder is determined by the existence of an adiabatic state -- a time-dependent eigenstate of the Hamiltonian \eqref{HN+1} --
 which connects the desired initial and final states $\left\vert 0_{1}0_{2}\ldots 0_{N}\right\rangle \left\vert 0\right\rangle$ and $\left\vert 1_{1}1_{2}\ldots 1_{N}\right\rangle \left\vert N\right\rangle $.
The energy of this adiabatic states is either below (for up-chirp, when the slope $\delta_0>0$) or above (for down-chirp, when the slope $\delta_0<0$) all other adiabatic energies. 
By imposing adiabatic evolution conditions, we can force the system to follow this adiabatic state, thereby arriving in the desired Fock state $\left\vert 1_{1}1_{2}\ldots 1_{N}\right\rangle \left\vert N\right\rangle $ in the end.

\begin{figure}[htb]
\includegraphics[width=\columnwidth]{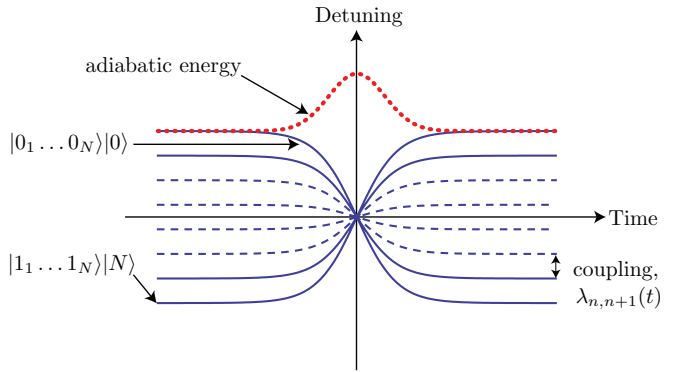}
\caption{(Color online.) The motional ground state and $N$-phonon Fock state of the ion chain can be connected adiabatically using a level-crossing, providing that the couplings are chosen to satisfy the adiabaticity condition, (\ref{adiabaticity}). To emphasise the `bow-tie' structure of the level-crossing, the energies of all of  the diabatic states (blue curves) are shifted by $-N\delta(t)/2$ from those in Eq. (\ref{HN+1}). The red dotted curve shows the particular adiabatic pathway used for Fock state preparation.
}
\label{level_crossing_scheme}
\end{figure}

To be specific, we choose to use a hyperbolic-secant pulse shape and a hyperbolic-tangent detuning,
\begin{subequations}\label{AE}
\begin{eqnarray}
\Omega (t) &=&\Omega _{0}\;\text{sech}\left( t/T\right) ,\label{Rabi} \\
\delta ( t) &=&\delta _{0}\tanh \left( t/T\right) .\label{DeltaAE}
\end{eqnarray}
\end{subequations} 
When only two states are involved, this model is exactly soluble analytically and is known as the complex-sech model in nuclear magnetic resonance \cite{complex-sech}
 and the Allen-Eberly-Hioe model in quantum optics \cite{AE};
 these are special cases of the more general Demkov-Kunike model \cite{DK}.
The model \eqref{AE} can therefore be viewed as an $(N+1)$-state generalization of this model.

There are several analytic solutions available in the literature for multistate bow-tie level-crossing problems.
Analytic bow-tie solutions have been found for three \cite{Carroll} and $N$ states \cite{Ostrovsky97,Harmin,Brundobler}, in which the coupling is constant and the detunings are linear in time.
However, in these solutions a single state is assumed to interact with all other states, which in turn do not interact with each other;
 this linkage is unfortunately different from the chainwise linkage in our Morris-Shore ladder.
A chainwise-coupled bow-tie linkage pattern occurs in an analytic model of an rf-pulse controlled Bose-Einstein condensate output coupler \cite{Mewes,VitanovBEC};
 here the underlying SU(2) symmetry allows for a reduction to an equivalent soluble two-state problem.
This symmetry, however, imposes strict relations between the couplings, which unfortunately are not met in our Morris-Shore ladder.

Compared to other types of level crossings, for instance the Landau-Zener model \cite{LZ},
 the sech-tanh model \eqref{AE} produces a relatively fast adiabatic passage, with a very  convenient and controllable excitation-profile, with a very high probability in a certain range around resonance and rapidly vanishing probability outside this range; the high-probability range is efficiently controlled by the chirp-rate $\delta_{0}$ \cite{RM}. 
A flat and narrow excitation profile is particularly relevant here, since it is important that our laser pulse does not excite unwanted vibrational sidebands, and this point is discussed further in Sec. \ref{LD}.

\subsubsection{Adiabatic conditions}

Because a multistate analytic solution for the model \eqref{AE} does not appear possible, we resort to numerical simulations;
 however, we shall make use of analytic insights derived from the above models in order to estimate the adiabaticity conditions.
A transfer efficiency of $1-\epsilon$ \mbox{(with $\epsilon\ll1$)} in the exactly soluble two-state case requires the coupling strength and the detuning to satisfy \cite{AE}
\begin{equation}\label{adiabaticity-2}
\frac{(\pi \eta \Omega_0T)^{2}}{2\ln(1/\epsilon)}\gtrsim \pi\delta _{0}T\gtrsim \ln(1/\epsilon), 
\end{equation}
where the signs of $\delta _{0}$ and $\Omega _{0}$ are unimportant and assumed positive for definiteness.
We have verified numerically that in the multistate Morris-Shore bow-tie ladder this condition is amended to
\begin{equation}\label{adiabaticity}
\frac{(\pi \eta \Omega_0T)^{2}}{2N\ln(1/\epsilon)}\gtrsim \pi\delta _{0}T\gtrsim N\ln(1/\epsilon).
\end{equation}
The important conclusion from here is that both the chirp rate $\delta_0$ and the peak Rabi frequency $\Omega_0$ have to increase linearly with the number of states $N$
 for the fidelity to remain above the value $1-\epsilon$.
We note that the linear dependence of $\Omega_0$ on $N$ agrees with that found in other systems involving multiple level crossings \cite{unanyan2002}
 and other types of multistate adiabatic passage \cite{vitanov1998}.

\subsubsection{Mechanism of Fock state creation}

The technique operates as follows. First, we apply a common laser pulse to all $N$ ions, tuned near to the blue sideband. 
By adiabatically sweeping the detuning $\delta \left( t\right) $ through resonance, all population will be transferred from the lowest to the highest energy state,
\footnote{In fact, the state also acquires a global phase factor. For brevity and clarity of expression, global phase factors are omitted in this section -- however, relative phases are included explicitly. 
We shall return to this point in more detail in subsection \ref{phases}.} 
\begin{equation}
\left\vert 0_{1}0_{2}\ldots 0_{N}\right\rangle \left\vert 0\right\rangle \rightarrow \left\vert 1_{1}1_{2}\ldots 1_{N}\right\rangle \left\vert N\right\rangle .  \label{FS}
\end{equation}

In an optional second step, to create number-states of higher phonon number, the laser is tuned near the carrier or red-sideband resonance, in order to return the population adiabatically, as follows:

(a) for the carrier transition 
\begin{equation}
\left\vert 1_{1}1_{2}\ldots 1_{N}\right\rangle \left\vert N\right\rangle
\rightarrow \left\vert 0_{1}0_{2}\ldots 0_{N}\right\rangle \left\vert N\right\rangle;
\label{SSC}
\end{equation}

(b) for the red-sideband transition 
\begin{equation}
\left\vert 1_{1}1_{2}\ldots 1_{N}\right\rangle \left\vert N\right\rangle
\rightarrow \left\vert 0_{1}0_{2}\ldots 0_{N}\right\rangle \left\vert 2N\right\rangle .
\label{SSB}
\end{equation}
Hence, by repeatedly applying the blue-carrier or blue-red transition cycles, it is possible to create the phonon Fock states $\left\vert kN\right\rangle $ (with $k=1,2,\ldots $). 

\begin{figure}[tbp]
\includegraphics[width=0.8\columnwidth]{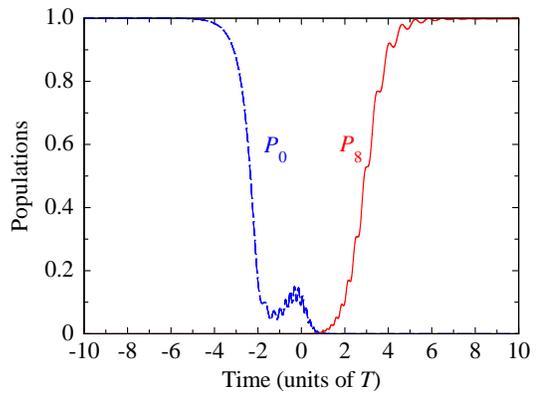}
\caption{(Color online.) Time evolution of the populations $P_0$ and $P_8$ 
which correspond to the chain containing zero and eight motional quanta,
for the case of \mbox{$N=8$} ions.
The Schr\"odinger equation was solved using the Hamiltonian given in Eq. (\ref{Hamiltonian_COM}), i.e. neither the Lamb-Dicke approximation, nor the vibrational rotating-wave approximation were applied. The time-dependences of the Rabi frequency $\Omega (t) $ and the detuning $\protect\delta (t)$ were of sech-tanh form  (\protect\ref{AE}), and the parameters were: $g(0)T=12.5$, $\delta _{0}T=10$, $\eta=0.3$, $g(0)=\nu/20\pi$. The final fidelity is $99.8\%$.}
\label{Fig-3}
\end{figure}

Figure \ref{Fig-3} shows the time evolution of the populations $P_0$ and $P_8$ of the states 
\mbox{$\left\vert0_{1}\dots0_{8}\right\rangle \left\vert 0\right\rangle$} and \mbox{$\left\vert 1_{1}\ldots1_{8}\right\rangle \left\vert 8\right\rangle$} for the case of eight ions. 
The initial ground-state population is completely transferred through
several intermediate entangled states to the highest-energy state which is an 
eight-phonon Fock state. We note that $\eta=0.3$ for the example given in Fig. 4. Despite this relatively high value of the Lamb-Dicke parameter, the technique still works with a high fidelity.

\begin{figure}
\includegraphics[width=0.75\columnwidth]{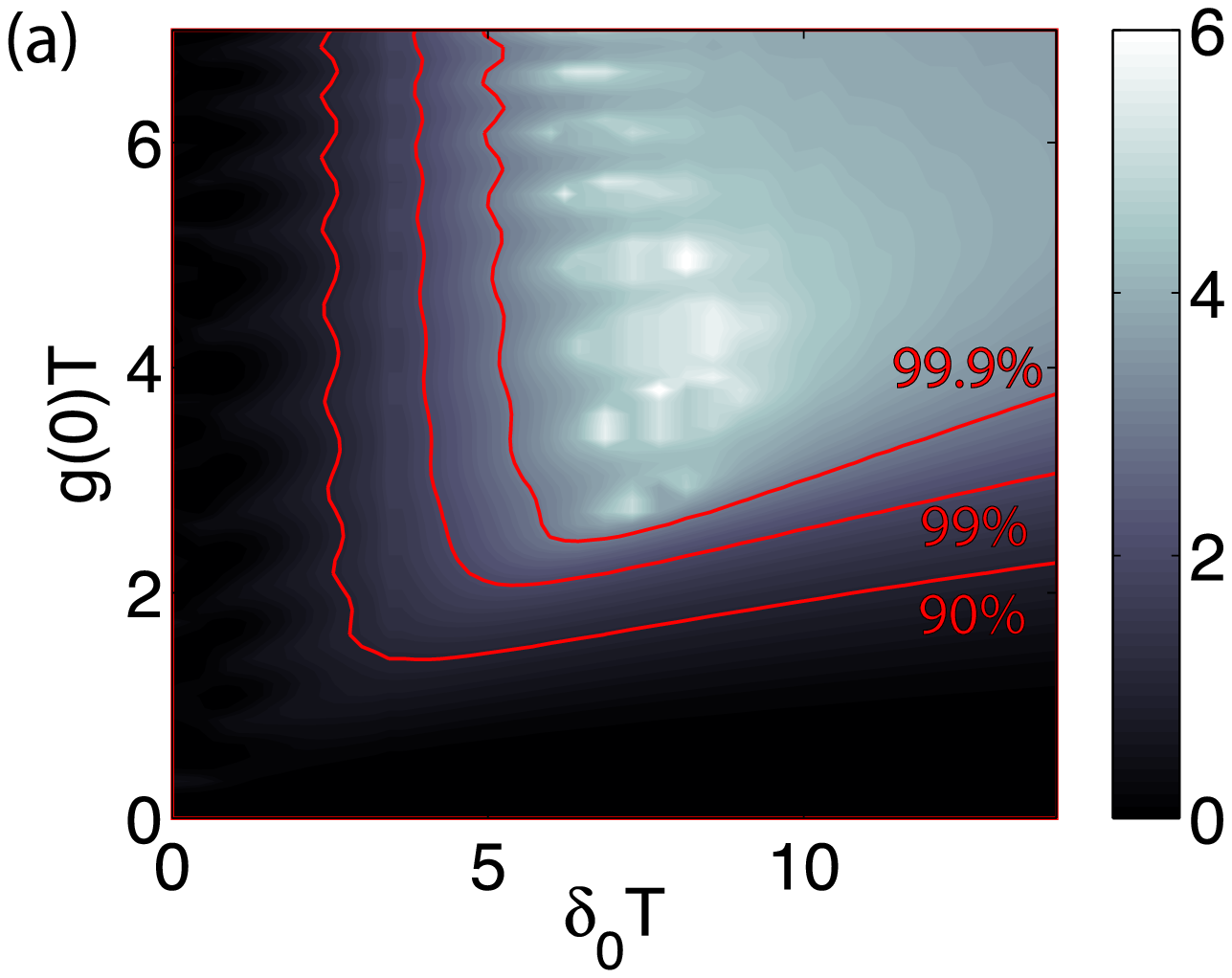}
\includegraphics[width=0.75\columnwidth]{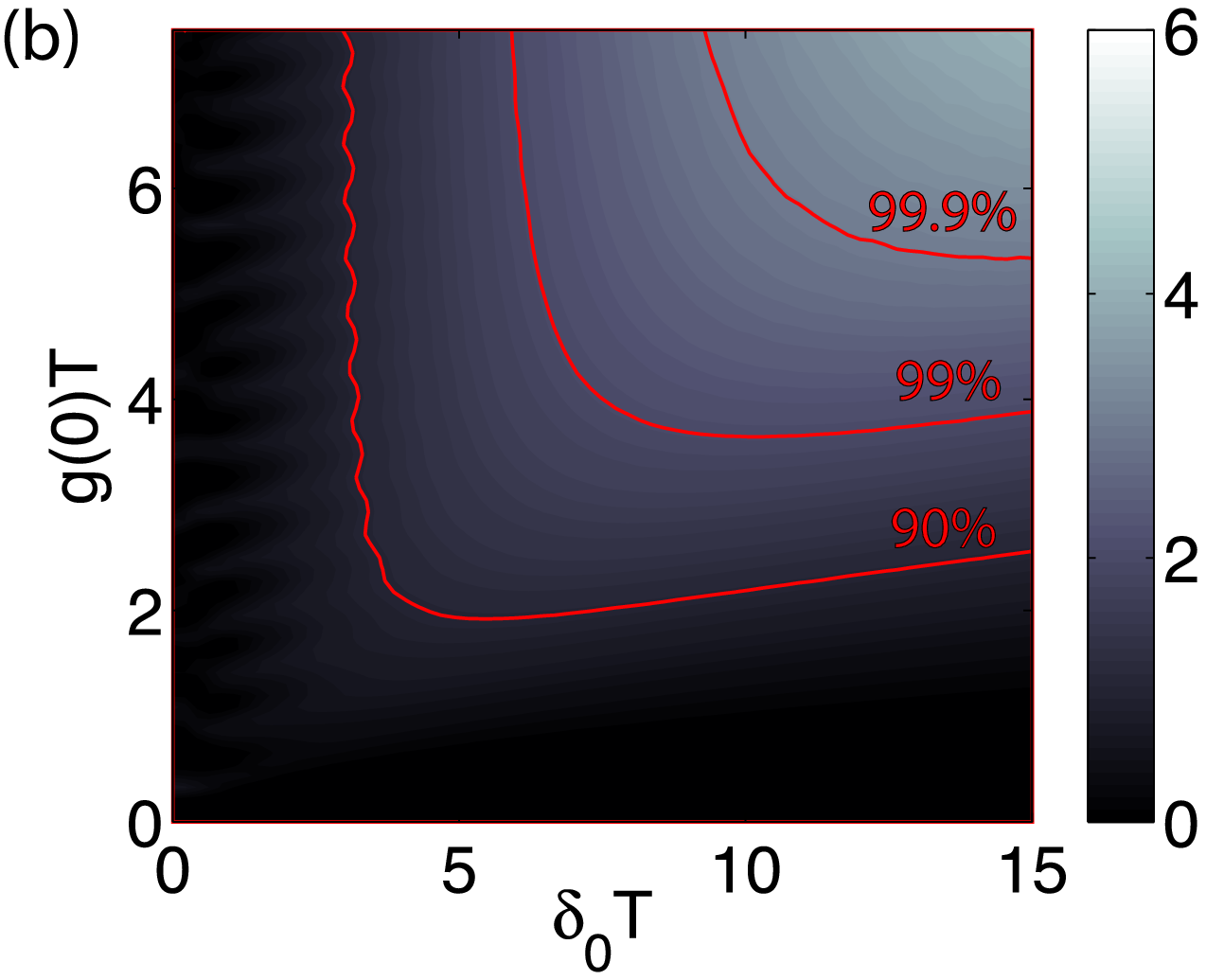}
\caption{(Color online.) Numerically calculated transfer efficiency for the creation of a five-phonon Fock state. The contour plot shows \mbox{$-\log(1-\mathcal{F})$}, (where $\mathcal{F}$ is the fidelity of Fock-state creation), plotted versus the scaled coupling $g(0)T$ and scaled detuning $\delta_{0}T$. The values of the Lamb-Dicke parameter are respectively: \mbox{(a) $\eta=0.1$}; \mbox{(b) $\eta=0.5$}. The simulations solve the Schr\"odinger equation with Hamiltonian (\ref{Hamiltonian_COM}) and neither the Lamb-Dicke approximation, nor the vibrational rotating-wave approximation is employed.}
\label{robust_plot}
\end{figure}

Since the technique described above uses adiabatic passage, the
transfer efficiency remains high even for considerable variations in the
laser intensity and detuning -- providing that the condition (\ref{adiabaticity}) is satisfied. To demonstrate this fact, Fig. \ref{robust_plot} shows a contour plot of the transfer efficiency as a function of $\delta_{0}T$ and $g(0)T$. 
The plot has a broad island of high transfer efficiency in the top-right-hand corner, and in this region the process is insensitive to fluctuations in the experimental parameters. This robustness represents an advantage over conventional techniques requiring precise pulse areas and frequencies.

\subsection{Creation of Fock states via a level crossing on the red-sideband}
\label{red_sideband} 
We note that the same Fock states can also be created in a single step using the red rather than blue motional sideband. 
In this case, the Hamiltonian is of Jaynes-Cummings, rather than anti-Jaynes-Cummings type
\bea
\mathbf{H}_{I}\left( t\right) = \hbar g\left( t\right) \sum_{j=1}^{N}\left[a\sigma _{j}^{+}e^{i\int_{t_{i}}^{t}\delta \left( \tau \right) d\tau -i\phi_{j} + i\pi/2} + h.c. \right]
\label{red_Hamiltonian}
\eea
which conserves the total number of excitations. The generalized
Morris-Shore transformation again applies \cite{LV} but in this case, the
Morris-Shore states in Eq. (\ref{MS_states}) are replaced by 
\begin{equation}
\mathbf{U}(t)\left\vert W_{N-n}^{N}\right\rangle \left\vert n\right\rangle 
\hspace{5mm}(n=0,\ldots ,N). 
\label{MS_states_red}
\end{equation}
Therefore an $N$-phonon Fock state can be created as follows. 
Firstly, the ions are initialized in their upper state, and the chain cooled to its ground state of motion, i.e. $|1_{1}1_{2}\ldots 1_{N}\rangle |0\rangle $. 
Subsequently, a common laser pulse is applied, and swept adiabatically through resonance with the red sideband. 
This transfers the system sequentially through the ladder of intermediate entangled states given in Eq. (\ref{MS_states_red}) and into the target state, $| 0_{1}0_{2}\ldots 0_{N}\rangle | N\rangle $. 
Higher-phonon states $|kN\rangle $ may be created by using the red-carrier or red-blue transition cycles in the manner outlined in subsection \ref{create_Fock}.

\subsection{Performance beyond the Lamb-Dicke limit}
\label{LD}
It is clear from Figs. 4 and 5 that while our technique is designed to operate in the Lamb-Dicke regime, it can tolerate significant values $\eta$ and still perform with high fidelity. This represents another clear advantage of the adiabatic technique presented in this article over conventional approaches to motional state engineering. To see why the technique can function with a range of $\eta$ values, we note that the terms present in Eq. (\ref{Hamiltonian_full}) which were discarded en-route to Eq. (\ref{Hamiltonian}) can be categorised into three groups, as follows:
(i)	resonant terms involving only operators from the centre-of-mass mode;
(ii)	non-resonant terms involving only operators from the centre-of-mass mode;
(iii)	non-resonant terms involving operators corresponding to higher-order axial normal modes.
%

The three groups of terms have the following effects.
Terms in category (i) simply adjust the strength of the Morris-Shore couplings between different levels in the longest chain. It is well-known that adiabatic techniques can tolerate significant variations in coupling strength and so, unsurprisingly, these terms do not harm the procedure. 
Category (ii) terms have a negligible effect providing that the vibrational rotating-wave approximation holds. In fact, application of the rotating-wave approximation is favoured both by a small Lamb-Dicke parameter, and also by the condition of adiabatic dynamics and these terms have a negligible effect even for $\Omega_{0}\lesssim\nu/\sqrt{N}$ in this regime.

The terms in category (iii) correspond to multi-phonon transitions, at sideband frequencies $(\pm\nu_{p} \pm \nu_{q}  + \nu)$ \footnote{These are the $\mathcal{O}(\eta^{2})$ terms; there are also higher-order terms, which are smaller again in magnitude.}, where $p$ and $q$ run over all vibrational modes other than the centre-of-mass mode \cite{buzek}; importantly, these are all non-resonant transitions. 
Because the sech-tanh pulse shape chosen in this paper is extremely selective, with an excitation profile that is both flat and narrow, excitation of off-resonant modes is strongly inhibited \cite{RM}.
In order to qualify this statement, we note that for the high-fidelity region in Fig. \ref{robust_plot}a [i.e. $g(0)T\sim6$, $\delta_{0}T\sim10$, $g(0)\sim\nu/20\pi$], the maximum pulse detuning is on the level of $2.6\%$ of the trap-frequency. By contrast, for a chain of ten ions, and taking into account the terms up to order $\eta^{2}$, we find that the motional-sideband with frequency closest to the sideband we address is detuned by approximately 11\% of the trap-frequency -- far enough from resonance to remain  unexcited, due to the high selectivity of the chosen sech-tanh pulse shape. Taking into account those terms proportional to $\eta^{3}$, the nearest motional sideband is still off-resonance by 4.7\% of the trap frequency, and also remains unexcited. Each successive group of terms also has a coupling strength that is of a factor of $\eta$ weaker again that the group before -- further aiding selectivity.
Therefore, from our numerical simulations, and also from the arguments outlined above, we conclude that our proposed technique should operate with high fidelity, even on the border of the Lamb-Dicke regime.

\section{Creation of motional Fock state superpositions and internal GHZ states}
\label{create_GHZ} 
\subsection{Creation of ionic GHZ states}

\begin{figure}
\includegraphics[width=0.95\columnwidth]{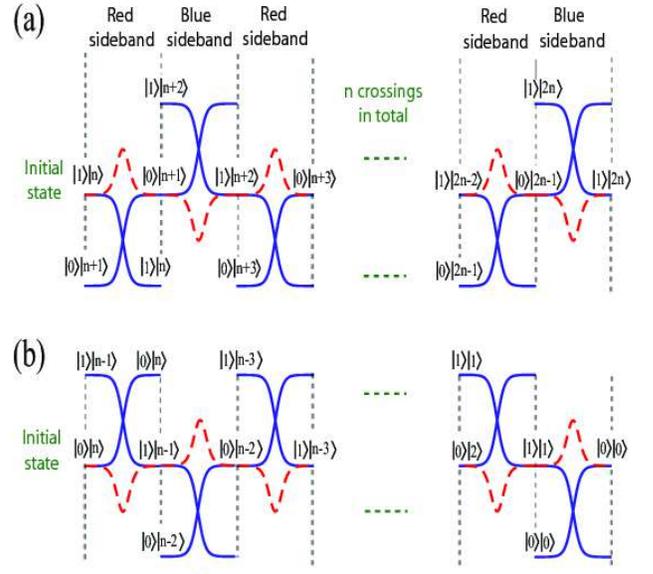}
\caption{(Color online.) Sequence of level crossings used to prepare the state (\ref{ground_sup}) for even $n$. 
The routes taken by the adiabatic states are shown as red dashed curves, while the bare energies are shown as solid blue curves. 
The initial state of the first ion is: (a) internal state $\vert 1\rangle$, (b) internal state $\vert 0\rangle$. 
After the $n$ crossings shown above, a final level-crossing on the red sideband has the effect \mbox{$\vert1\rangle\vert2n\rangle\rightarrow\vert0\rangle\vert2n+1\rangle$},
 while leaving the state \mbox{$\vert 0\rangle\vert 0\rangle$} unchanged.}
\label{level_crossings}
\end{figure}

A modified version of the above technique can be used to create generalized GHZ states \cite{leibfried2005,GHZ}. 
This is done in a two-stage process. 
First the number-state superposition 
\begin{equation}
\left\vert 0_{1}0_{2}\ldots 0_{2n+1}\right\rangle \left( c_{0}\left\vert 0\right\rangle +c_{1}\left\vert 2n+1\right\rangle \right)
\label{ground_sup}
\end{equation}
is prepared, and then the motional excitations are mapped onto the internal state of the ion chain.
The state (\ref{ground_sup}) can be created as follows. 
Initially we prepare the system in the product state $\left\vert 0_{1}0_{2}\ldots 0_{2n+1}\right\rangle \left\vert n\right\rangle $ using the technique described above,
 and then perform a controlled rotation of the state of the first ion using the carrier-frequency transition, to obtain
\begin{equation}
\left( c_{0}\left\vert 0_{1}\right\rangle +c_{1}\left\vert 1_{1}\right\rangle \right) \left\vert 0_{2}\ldots 0_{2n+1}\right\rangle\left\vert n\right\rangle
 \label{step_one}
\end{equation}
for even $n$ or
\begin{equation}
\left( c_{1}\left\vert 0_{1}\right\rangle +c_{0}\left\vert 1_{1}\right\rangle \right) \left\vert 0_{2}\ldots 0_{2n+1}\right\rangle\left\vert n\right\rangle  
\label{step_oneb}
\end{equation}
for odd $n$.
Next, a sequence of avoided level crossings is induced by addressing the first ion, with pulses alternating between the first red- and blue-sidebands (ending with the red sideband). 
We note that such level crossings have been performed experimentally by \mbox{Wunderlich et al}, each with a fidelity above $99\%$ \cite{Wunderlich}. 
During each avoided crossing, a phonon is removed or added, depending on the internal state of the first ion, and after $(n+1)$ crossings, the state (\ref{ground_sup}) is reached. 
This sequence of avoided level crossings is shown schematically in Fig. \ref{level_crossings}.

In the return step, all $(2n+1)$ ions are addressed simultaneously by a laser beam swept through the red-sideband transition in the manner described in Sec. \ref{red_sideband}. 
This maps the phonons onto ionic excitations and gives
\begin{equation}
\left\vert 0_{1}0_{2}\ldots 0_{2n+1}\right\rangle \left( c_{0}\left\vert 0\right\rangle +c_{1}\left\vert 2n+1\right\rangle \right) \rightarrow
\left\vert \text{GHZ}\right\rangle \left\vert 0\right\rangle ,  \label{GHZ}
\end{equation}
where 
\begin{equation}
\left\vert \text{GHZ}\right\rangle =c_{0}\left\vert 0_{1}0_{2}\ldots
0_{2n+1}\right\rangle +c_{1}e^{i\xi }\left\vert 1_{1}1_{2}\ldots
1_{2n+1}\right\rangle  \label{GHZ1}
\end{equation}
is a generalized GHZ state with arbitrary amplitudes, and the known phase $\xi$ is specified in Eq. (\ref{xi_def}) below. The above technique creates GHZ states of odd numbers of ions, from which even-numbered GHZ states can be derived using only single-ion rotations and measurement.

\subsection{Creation of superpositions of motional Fock states}

From the preceding analysis, it is clear that our technique can also be used to convert entanglement between the ions' internal states into a coherent superposition of two different states of their collective motion. 
Beginning in a generalized $N$-ion GHZ state and applying a common pulse to the whole chain (swept through the red-sideband transition) gives
\begin{equation}
\left\vert \text{GHZ}\right\rangle \left\vert 0\right\rangle \rightarrow
\left\vert 0_{1}0_{2}\ldots 0_{N}\right\rangle \left( c_{0}\left\vert 0\right\rangle +c_{1}e^{i\xi}\left\vert N\right\rangle \right),
\label{motional_sup}
\end{equation}
and applying the blue-carrier transition cycle $k$ further times creates the
following superposition state 
\begin{equation}
\left\vert \Psi _{k}\right\rangle =\left\vert 0_{1}0_{2}\ldots
0_{N}\right\rangle \left(c_{0}\left\vert k
N\right\rangle +c_{1}e^{i\xi'}\left\vert (k+1) N\right\rangle \right).
\label{higher_motional_sup}
\end{equation}

\subsection{Phase factors} \label{phases}

In general, the phase factor appearing in Eqs. (\ref{GHZ1}), (\ref{motional_sup}) and (\ref{higher_motional_sup}) has two components,
\begin{equation}
\xi =\frac{1}{\hbar }\int_{t_{i}}^{t_{f}}E(t^{\prime })dt^{\prime
}-\sum_{j=1}^{N}\phi _{j}\langle1_{j}\vert\big(\vert\psi _{f}\rangle -\vert\psi
_{i}\rangle \big),  
\label{xi_def}
\end{equation}
where $E(t)$ is the eigenvalue of the Hamiltonian (\ref{HN+1}) corresponding to the state used for the adiabatic transfer process, and $\phi _{j}$ is given in Eq. (\ref{phi_def}). 
In situations where the whole state accumulates a global phase, $\xi $ has been omitted above,
 but in cases where only a constituent part of the total state acquires this phase (i.e. in Sec.~\ref{create_GHZ}), it has been included explicitly.

The dynamical part of $\xi$ [the integral in Eq.~(\ref{xi_def})] can be controlled by choosing appropriate values for the area and detuning of the laser pulse. 
Alternatively, if a Raman-coupled hyperfine qubit encoding is chosen, then the adiabatic passage described in section \ref{Sec-DP} can be performed using a multi-ion dark state. 
In this case, $E(t)$ is identically zero throughout the transfer process and the state acquires no dynamical phase. 
Full details of this dark-state technique will be described elsewhere \cite{LV2}.

The second term in Eq.~(\ref{xi_def}) arises because the ions are not equally spaced along the trap axis. 
This means that using a single laser pulse, it is impossible to set all of the individual phase factors $e^{i\phi _{j}}$ to unity.
However, by symmetry, it \emph{is possible} to set $\sum_{j=1}^{N}\phi _{j}=0$. 
Hence when the whole chain is addressed simultaneously, this second term in (\ref{xi_def}) always vanishes. 
For individual addressing of each ion, this term can always be chosen to be zero, even when only part of the chain is used.

\section{Heating effects}\label{heating} 

\begin{figure}
\includegraphics[width=0.95\columnwidth]{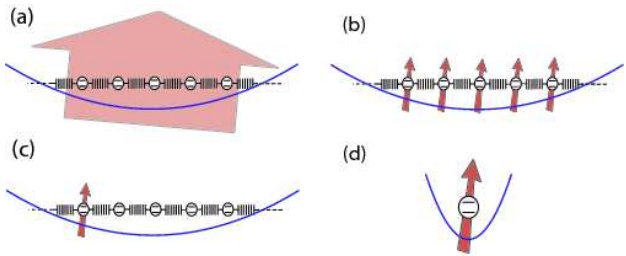}
\caption{(Color online.) Alternative experimental routes to the creation of vibrational number states using adiabatic passage:
 (a) Common addressing of all ions;
 (b) Individual addressing of each ion;
 (c) Repeated addressing of a specific ion in the chain; (d) Repeated addressing of a single trapped ion.
Approaches (a) and (d) require that the centre of mass mode is used, whereas (b) and (c) may be implemented using higher-order vibrational modes.}
\label{three_approaches}
\end{figure}

Vibrational heating is one of the major limiting factors in ion-trap quantum information.
Here we estimate the heating effects in the proposed global addressing technique and make a comparison with three alternative adiabatic-passage techniques:
 one that uses $N$ ions and local addressing (which is a variation of our technique and allows us to use higher-order phonon modes), and two variations on the traditional one that uses a single ion.
These different implementations are sketched schematically in Fig. \ref{three_approaches}. 

\subsubsection{Center-of-mass mode and global addressing}
\label{global_heating}
In writing Eq.~(\ref{Hamiltonian}), we have assumed that the coupling $g(t)$ between the internal and motional states is the same for each ion,
and produced by a single laser pulse, tuned close to the blue sideband of the centre-of-mass mode.
While using a single laser pulse makes the technique extremely simple to implement,
 the heating rates for the centre-of-mass mode are typically larger than for higher-order vibrational modes \cite{James1998b, King}.

As follows from the adiabatic condition \eqref{adiabaticity}, the required pulse area $A=\pi\Omega_0 T$ of a sech-tanh pulse for fidelity $1-\epsilon$ scales in proportion to $N$ and $\ln(1/\epsilon)$.
For a fixed peak Rabi frequency $\Omega_0$, the characteristic pulse width for our global-addressing technique is therefore
\be\label{T_global}
 T = \frac{\sqrt{2}\;N\ln(1/\epsilon)}{\pi\eta\Omega_0} = \frac{\sqrt{N}\;\ln(1/\epsilon)}{\sqrt{2}\;\pi g(0)}.
\ee
The total interaction time for a sech pulse can be taken as $T_{\text{global}}\approx 10T$.

In order to estimate deleterious effects of heating on our scheme, we assume $g(0)\approx \nu/20\pi$.
This is the reasonable maximal value of the coupling (implying minimal pulse duration), for which 
 no extraneous excitations in other vibrational modes are created \cite{James} and the rotating-wave approximation is satisfied
\footnote{The rotating-wave approximation is easy to satisfy in the adiabatic limit, since any transitions driven off-resonantly on the carrier transition cancel out over the course of the pulse. 
Therefore, $\Omega\lesssim\nu/\sqrt{N}$ is enough to satisfy the rotating-wave approximation in this case (see e.g. \cite{Wunderlich}).}. 
For a trap frequency of \mbox{$\nu/2\pi=4$MHz} as in \cite{schmidtkaler2003}, we find that total interaction time of $T_{\text{global}}\approx 150\ \mu$s
 is sufficient for 0.9999 fidelity ($\epsilon=10^{-4}$) for $N=8$ ions.
With a heating rate per ion of approximately 5 phonon/sec \cite{schmidtkaler2003},
 we predict heating of about $6\times10^{-3}$ phonons in total. 
\subsubsection{Higher vibrational modes and local addressing}

Decoherence effects may be further reduced by using a higher-order vibrational mode with a lower heating rate, if each ion is addressed individually, i.e., 
\begin{align}
\mathbf{H}_{I}\left( t\right) = \hbar \sum_{j=1}^{N}g_{j}\left(t\right) \Big[ a^{\dagger }\sigma_{j}^{+}e^{i\int_{t_{i}}^{t}\delta_{j}(\tau)d\tau -i\phi _{j}+i\pi/2}  +h.c. \Big].
\end{align}
Each individual coupling strength,
\begin{equation}
g_{j}(t)=\frac{\eta\Omega_{j}(t)\kappa_{p}^{j}}{2\sqrt{N}}, 
\end{equation}
now contains the dimensionless factor $\kappa_{p}^{j}$ defined in Eq. (\ref{Hamiltonian_full}) \cite{buzek}. By tuning the relative amplitude of each individual laser appropriately, it is possible to produce the Hamiltonian (\ref{Hamiltonian}). 

\subsubsection{Using a single ion}

Alternatively, as a trade-off between the technical simplicity of using a single laser pulse and the low heating rates afforded by higher-order modes,
 one can use a single ion repeatedly instead of $N$ ions in a single step;
 then phonons can be added or subtracted sequentially instead of all in one step. In this case, the difference between addressing one out of a chain of $N$ ions (Fig. \ref{three_approaches}c) and a single trapped ion (Fig. \ref{three_approaches}d) is the heating rate, which scales linearly with $N$. Below, for clarity, we consider the approach involving one ion from a chain of $N$ ions (Fig. \ref{three_approaches}c), and compare this to the global addressing technique discussed in subsection \ref{global_heating} (Fig. \ref{three_approaches}a).

The advantages of using a single ion are the lower heating rates of higher-order modes, and the use of a single laser. It also permits the creation of number states of arbitrary phonon number, rather than in multiples of the total number of ions. 
However, we stress that the single-ion multiple-step scenario requires a significantly higher fidelity in each step, $1-\epsilon/N$ instead of $1-\epsilon$, for the global-addressing technique presented here.
This higher fidelity requires larger pulse duration for a fixed peak Rabi frequency.

The characteristic pulse width for a repeated-local-addressing technique with a single ion requires $N$ steps with an $N$ times lower error [cf. Eq.~\eqref{T_global}],
\be\label{T_local}
T = \frac{\sqrt{2}\;N\ln(N/\epsilon)}{\pi\eta\Omega_0} = \frac{\sqrt{N}\;\ln(N/\epsilon)}{\sqrt{2}\;\pi g(0)}.
\ee
The total interaction time for a sech pulse is again $T_{\text{local}}\approx 10T$.
The single-ion implementation therefore requires longer time than the global-addressing technique, by a factor of $\ln(N/\epsilon)/\ln(1/\epsilon)$.
For example, for $\epsilon=0.01$ and $N=8$ ions, this factor is about 1.5.
Therefore the number of heating events during the preparation of the Fock states increases by the same factor.

In summary, our global-addressing technique allows for a shorter time duration than single-ion techniques, and in addition, offers an appealing simplicity of implementation, with the need to engineer a single interaction only.

Finally, we consider the preparation of motional states of a single ion, as shown in Fig. \ref{three_approaches}d, whereby a sequence of adiabatic pulses can be applied to add or remove phonons one at a time. In this respect the proposal is similar to earlier work involving single ions, such as \cite{Meekhof} which used sequential pulses of precise area. However, our method possesses a significant advantage over previous work -- namely a natural robustness with respect to experimental imperfections. This robustness arises from the adiabaticity of the process, as discussed in Sec.~\ref{section_Fock}, and should allow a significant increase in the fidelity of the preparation procedure.

\section{Conclusions}
\label{conclusions} 

We have shown how vibrational Fock states of a trapped-ion chain can be produced in a single step and using a single laser pulse. 
This is achieved by sweeping the laser frequency through resonance with a motional-sideband of the centre-of-mass vibrational mode. 
The key theoretical ingredient of our proposal is a particular choice of basis states in which to analyse the problem (the generalized Morris-Shore basis),
 wherein the Hilbert space factorizes into a collection of independent ladders.
Sweeping the laser detuning through resonance induces a `bow-tie' multiple-level crossing.
With the proviso that the pulse area and detuning satisfy an adiabaticity condition, this pulse connects the lowest and highest levels of each ladder. 
Moreover, the zero-phonon and $N$-phonon Fock states are located at the opposite ends of the longest chain, which provides an extremely simple recipe for creating phonon Fock states in a single step. 
Since the state preparation takes place adiabatically, it is naturally robust against fluctuations in the experimental parameters, such as the laser detuning and chirp, the Rabi frequency,
 and the pulse duration. Numerical simulations show that the technique also performs well on the boundary of the Lamb-Dicke regime.

Our technique may also be used to prepare GHZ states of the ions' internal degrees of freedom and superpositions of motional Fock states of the whole chain. 
We estimate that the fidelity of each step could be over $99\%$ for an eight-phonon Fock state if the centre-of-mass mode and a single laser pulse are used,
 and we have also suggested several ways in which heating effects could be further reduced.

\acknowledgments

This work has been supported by the EU ToK project CAMEL (Grant No. MTKD-CT-2004-014427), the EU RTN project EMALI (Grant No. MRTN-CT-2006-035369),
 the EU Integrated Project QAP (Grant No. IST-015848), the Royal Society, 
and the Bulgarian National Science Fund Grants No. VU-205/06 and No. VU-301/07.


\end{document}